# Part probabilistic cloning of linearly dependent states

Pinshu Rui[1,2], Wen Zhang[1*], Yanlin Liao and Ziyun Zhang[1]

1 Key Laboratory of Optoelectronic Information Acquisition & Manipulation of Ministry of Education of China, School of Physics & Material Science, Anhui University, Hefei 230601, China

2 Anhui Xinhua University, Hefei 230088, China

**Abstract**  From Ref. [Phys. Rev. Lett. 80(1998)4999] one knows that the quantum states secretly chosen from a certain set can be probabilistically cloned with positive cloning efficiencies if and only if all the states in the set are linearly independent. In this paper, we focus on the probabilistic quantum cloning (PQC) of linearly dependent states with nonnegative cloning efficiencies. We show that a linearly independent subset of the linearly dependent quantum states $\{|\Psi_1\rangle, |\Psi_2\rangle, \cdots, |\Psi_n\rangle\}$ can be probabilistically cloned if and only if any one state in the subset can not be expressed as the linear superposition of the other states in the set $\{|\Psi_1\rangle, |\Psi_2\rangle, \cdots, |\Psi_n\rangle\}$. The optimal possible cloning efficiencies are also investigated.

## I   INTRODUCTION

Because of the quantum no-cloning theorem[1], we know that an unknown quantum state can not be copied perfectly. However, it may be copied in an approximate or probabilistic way. The copy processes are known as the approximate quantum cloning[2-6] and the probabilistic quantum cloning (PQC)[7], respectively. PQC, initially presented by Duan and Guo in 1998[7], is used for probabilistically cloning the quantum states secretly chosen from a certain states set. It has been proved that, any quantum states secretly chosen from a certain states set can be probabilistically cloned if and only if all the states in the set are linearly independent[7]. In PQC, failing probabilities exist unless the states in the set are orthogonal to each other. However, the quantum state can be copied with the unit fidelity.

PQC has important applications in computation[8]. Moreover, it can be used for eavesdropping on the quantum cryptography. Therefore, after Duan and Guo's original work, PQC has attracted much attention[9-27]. The optimal PQC efficiencies for $M \to N$ ($M>N$) cloning of $m$ ($m \geqslant 2$) states have been given[9,10]. The simple cloning equations, the explicit unitary operations[11-16] and the quantum logic network[17,18] for PQC have been fully investigated. Some realistic proposals based on different physical systems have been put forward for $1 \to 2$ PQC of two states set[19,20]. Experimentally, PQC has been demonstrated in NMR system in 2011[21]. Further, the novel PQC[22,23], which produce the linear superposition of multiple copies of the input state, and the probabilistic cloning with supplementary information [24,25] have been studied.

---

* Corresponding author: wenzhang@ahu.edu.cn

PQC process contains an appropriate unitary operation and a projection measurement. The to-be-cloned state is secretly chosen from a certain states set. Therefore all the unitary operations corresponding to different to-be-cloned states in the set should be completely the same. The precondition of the existing works about PQC is that all the to-be-cloned states are cloned with positive real efficiencies. Just because of this prerequisite, linearly dependent states can not be probabilistically cloned. Then we naturally consider that, if the prerequisite of positive cloning efficiencies is discarded can a subset of linearly dependent states be probabilistically cloned? That is, some of the cloning efficiencies are positive while the other cloning efficiencies are zero. In this paper, we investigate the part PQC of linearly dependent states set and the optimal cloning efficiencies.

The rest of this paper is planed as follows. In Section II, we briefly review Duan and Guo's original work about PQC. In Section III, we investigate the part PQC of linearly dependent states. Finally, we extend the part PQC to the part discrimination of linearly dependent states and give a summary in section IV.

## II  DUAN-GUO'S ORIGINAL WORK ABOUT PQC[7]

Suppose the to-be-cloned state is randomly chosen from the states set $\{|\Psi_1\rangle, |\Psi_2\rangle, \cdots, |\Psi_m\rangle\}$. The 1→2 PQC of these states is expressed as

$$U|\Psi_j\rangle_x |\Sigma\rangle_y |P^{(0)}\rangle_z = \sqrt{\gamma_j} |\Psi_j\rangle_x |\Psi_j\rangle_y |P^{(j)}\rangle_z + \sqrt{1-\gamma_j} |\Phi^{(j)}\rangle_{xyz}, \quad (j=1,2,\cdots m), \qquad (1)$$

where $U$ is an appropriate unitary operation performed on the systems $x$, $y$ and $z$, $|\Sigma\rangle$ is the initial state of the system $y$, $|P^{(0)}\rangle$ and $|P^{(j)}\rangle$ are normalized states (not generally orthogonal), $|\Phi^{(1)}\rangle$, ..., $|\Phi^{(n)}\rangle$ are $m$ normalized states of the composite system $xyz$ (not generally orthogonal) and $\langle P^{(j)} | \Phi^{(k)} \rangle = 0$. The subspace spanned by the states $|P^{(1)}\rangle, \cdots, |P^{(m)}\rangle$ is denoted by the symbol $H_p$. During the cloning process, after the unitary evolution $U$ a projection measurement is carried out on the system $z$. With the positive cloning efficiency $\gamma_j$, the measurement projects the state of $z$ into the subspace $H_p$. Then the state of $x$ and $y$ collapses to $|\Psi_j\rangle_x |\Psi_j\rangle_y$ and the cloning is realized successfully. If we get other measurement projections with the probability of $1-\gamma_j$, the cloning process fails.

It has been proved that the necessary and sufficient condition for PQC of the states set

$\{|\Psi_1\rangle, |\Psi_2\rangle, \cdots, |\Psi_m\rangle\}$ with positive cloning efficiencies is all the states in the set are linearly independent (see Theorem 1 in Ref.[7]). Moreover, the theorem 2 in Ref.[7] shows that the legitimate cloning efficiencies of PQC can be obtained from the positive semidefinite of the matrix $X^{(1)} - \sqrt{\Gamma} X_z^{(2)} \sqrt{\Gamma^+}$, where $X^{(1)} = [\langle \Psi_j | \Psi_k \rangle]$, $X_z^{(2)} = [\langle \Psi_j | \Psi_k \rangle^2 \langle P^{(j)} | P^{(k)} \rangle]$ and $\Gamma = \Gamma^+ = \text{diag}(\gamma_1, \gamma_2, \cdots, \gamma_m)$. Using the theorem, it has been shown that the cloning efficiencies for cloning the two states set $\{|\Psi_1\rangle, |\Psi_2\rangle\}$ satisfies

$$\frac{\gamma_1 + \gamma_2}{2} \leq \frac{1}{1 + |\langle \Psi_1 | \Psi_2 \rangle|}. \qquad (2)$$

The equality in the inequality (2) holds if and only if $\gamma_1 = \gamma_2$ is satisfied.

From the review of Eq.[7] and the other references about PQC it can be seen that all the cloning efficiencies are assumed to be positive real numbers. It should be pointed out that, if some of the cloning efficiencies are zero and the other cloning efficiencies are small enough $X^{(1)} - \sqrt{\Gamma} X_z^{(2)} \sqrt{\Gamma^+}$ can also be positive semidefinite matrix. In this case, some of the states in the set can be probabilistically cloned while PQC of the other states always fails. This is the so-called part PQC. In the next section, we will study the part PQC of linearly dependent states.

### III  PART PQC OF LINEARLY DEPENDENT STATES

In this section, we study the part PQC of the linearly dependent quantum states set $S = \{S_m, S'_m\}$, where $S_m = \{|\Psi_1\rangle, |\Psi_2\rangle, \cdots, |\Psi_m\rangle\}$ is the maximal linearly independent subset of $S$ and $S'_m = \{|\Psi_{m+1}\rangle, |\Psi_{m+2}\rangle, \cdots, |\Psi_n\rangle\}$. The state in the subset $S'_m$ can be written as

$$|\Psi_J\rangle = \sum_{j=1}^{m} a_j^{(J)} |\Psi_j\rangle, \qquad (J = m+1, \cdots, n). \qquad (3)$$

Firstly, let us assume that the states in the set $S_m$ undergo the evolutions in Eq. (1) and all the cloning efficiencies are positive numbers ($\gamma_j > 0$). From Section II we know that the unitary operation $U$ in Eq. (1) exists because the states $|\Psi_1\rangle, |\Psi_2\rangle, \cdots, |\Psi_m\rangle$ are linearly independent. Then for the states in the subset $S'_m$, the evolutions are

$$U |\Psi_J\rangle_x |\Sigma\rangle_y |P^{(0)}\rangle_z = U \sum_{j=1}^{m} a_j^{(J)} |\Psi_j\rangle_x |\Sigma\rangle_y |P^{(0)}\rangle_z$$

$$= \sqrt{\gamma_J} \sum_k \sum_{j=1}^{m} b_{jk}^{(J)} |\Psi_j\rangle_x |\Psi_j\rangle_y |p_k\rangle_z + \sum_{j=1}^{m} a_j^{(J)} \sqrt{1-\gamma_j} |\Phi^{(j)}\rangle_{xyz}, \qquad (J = m+1, \cdots, n), \qquad (4)$$

where $|p_k\rangle$'s are orthogonal bases of the subspace $H_p$ and

$$\sqrt{\gamma_J}\sum_k\sum_{j=1}^m b_{jk}^{(J)}|\Psi_j\rangle_x|\Psi_j\rangle_y|p_k\rangle_z = \sum_{j=1}^m a_j^{(J)}\sqrt{\gamma_j}|\Psi_j\rangle_x|\Psi_j\rangle_y|P^{(j)}\rangle_z. \qquad (5)$$

From Eq. (4) one can see that, corresponding to the projection $|p_k\rangle_z$ the systems $x$ and $y$ collapse to the normalized state $\sum_{j=1}^m b_{jk}^{(J)}|\Psi_j\rangle_x|\Psi_j\rangle_y$. Because $\gamma_j > 0$ and the states in $S'$ are linearly independent, for any $j \in [1,\cdots,m]$ and $J \in [m+1,\cdots,n]$ we have

$$\sum_{j=1}^m b_{jk}^{(J)}|\Psi_j\rangle_x|\Psi_j\rangle_y \neq \sum_{j,j'=1}^m a_j^{(J)}a_{j'}^{(J)}|\Psi_j\rangle_x|\Psi_{j'}\rangle_y = |\Psi_J\rangle_x|\Psi_J\rangle_y. \qquad (6)$$

Eqs. (4,6) show that corresponding to the input state $|\Psi_J\rangle$ the projection $|p_k\rangle_z$ can not leads to the collapse of $|\Psi_J\rangle_x|\Psi_J\rangle_y$ and the cloning fails. However, the same projection results in the collapse of $|\Psi_j\rangle_x|\Psi_j\rangle_y$ for the input state $|\Psi_j\rangle$ [see Eq. (1)]. That is, from the projection $|p_k\rangle_z$ we do not know the PQC is successful or not. Consequently, linearly dependent states can not be probabilistically cloned.

From the previous demonstration one knows that, the projection of the subspace $H_p$ must cause the successful cloning in a legitimate PQC process. In the following, let us assume $\gamma_j \geq 0$. From Eqs. (4-6) we must let $\gamma_J = 0$ and consequently we have

$$\sum_{j=1}^m a_j^{(J)}\sqrt{\gamma_j}|\Psi_j\rangle_x|\Psi_j\rangle_y|P^{(j)}\rangle_z = 0. \qquad (7)$$

Because $|\Psi_1\rangle, |\Psi_2\rangle, \cdots, |\Psi_m\rangle$ are linearly independent, $\{|\Psi_j\rangle_x|\Psi_j\rangle_y|P^{(j)}\rangle_z, j=1,2,\cdots m\}$ is also a linearly independent states set. Then from Eq.(7) one knows that

$$a_j^{(J)}\sqrt{\gamma_j} = 0, \qquad (j=1,2,\cdots m, J=m+1,\cdots,n). \qquad (8)$$

If $A_j = \sum_{J=m+1}^n |a_j^{(J)}| = 0$, from Eq. (8) we know that $\gamma_j$ can be chosen as a real number and the state $|\Psi_j\rangle$ can be probabilistically cloned with the cloning efficiency $\gamma_j$. On the contrary, if $A_j \neq 0$, we have $\gamma_j = 0$ and $|\Psi_j\rangle$ can not be probabilistically cloned. Therefore the necessary and sufficient condition for part PQC of the states set $S$ is that some of $A_j$'s ($j=0,1,\cdots,m$) are zero. Combining with Eq. (3) we thus get the following theorem.

***Theorem* 1** — Part PQC of the states set *S* can be implemented by an appropriate unitary operation and a projection measurement if and only if there exists a linearly independent subset in which any one state can not be expressed as the linear superposition of the other states in *S*.

In the following, we derive the optimal cloning efficiencies $\gamma_j$ able to be attained by a part PQC. The to-be-cloned states are secretly chosen from the set *S* which satisfies the Theorem 1. Without loss of generality, we express the linearly dependent states as $S = \{S_l, S'_l\} = \{S_m, S'_m\}$ where $S_l = \{|\Psi_1\rangle, |\Psi_2\rangle, \cdots, |\Psi_l\rangle\}$, $S'_l = \{|\Psi_{l+1}\rangle, |\Psi_{l+2}\rangle, \cdots, |\Psi_n\rangle\}$ and

$$|\Psi_J\rangle = \sum_{j=l+1}^{m} a_j^{(J)} |\Psi_j\rangle, \qquad (J = m+1, \cdots, n). \tag{9}$$

Obviously, $l < m$ and the states in the set $S_l$ can not be expressed as the linear superposition of the states $|\Psi_{l+1}\rangle$, $|\Psi_{l+2}\rangle, \cdots, |\Psi_n\rangle$. Similar to that in Eq. (1), the part PQC process is expressed as

$$U|\Psi_j\rangle_x |\Sigma\rangle_y |P^{(0)}\rangle_z = \sqrt{\gamma_j} |\Psi_j\rangle_x |\Psi_j\rangle_y |P^{(j)}\rangle_z + \sqrt{1-\gamma_j} |\Phi^{(j)}\rangle_{xyz}, \quad (j = 1, 2, \cdots n), \tag{10}$$

where $\gamma_j \geq 0$, $\prod_{j=1}^{l} \gamma_j > 0$ and $\gamma_{l+1} = \gamma_{l+2} = \cdots = \gamma_n = 0$. That is, states in the subset $S_l$ can be cloned with positive efficiencies while the cloning efficiencies for other states are zero. As a matter of fact, if $\gamma_{l+1} = \gamma_{l+2} = \cdots = \gamma_m = 0$, from Eqs. (9,10) we directly have

$$U|\Psi_J\rangle_x |\Sigma\rangle_y |P^{(0)}\rangle_z = \sum_{j=l+1}^{m} a_j^{(J)} |\Phi^{(j)}\rangle_{xyz} = |\Phi^{(J)}\rangle_{xyz}, \quad (J = m+1, \cdots n). \tag{11}$$

That is, from $\gamma_{l+1} = \gamma_{l+2} = \cdots = \gamma_m = 0$ one can get $\gamma_{m+1} = \gamma_{m+2} = \cdots = \gamma_n = 0$. Therefore in the following we only consider the PQC of the maximal linearly independent subset $S'$ with the assumption of $\gamma_{l+1} = \gamma_{l+2} = \cdots = \gamma_m = 0$. Like that in Ref. [7], we further assume $\sqrt{1-\gamma_j}|\Phi^{(j)}\rangle = \sum_{k=1}^{m} c_{jk} |\Phi^{(jk)}\rangle$, where $j = 0, 1, \cdots, m$ and $|\Phi^{(j1)}\rangle, |\Phi^{(j2)}\rangle, \cdots, |\Phi^{(jm)}\rangle$ are *m* orthonormal states of the composite system *xyz*. Then we have

$$U|\Psi_j\rangle_x |\Sigma\rangle_y |P^{(0)}\rangle_z = \sqrt{\gamma_j} |\Psi_j\rangle_x |\Psi_j\rangle_y |P^{(j)}\rangle_z + \sum_{k=1}^{m} c_{jk} |\Phi^{(jk)}\rangle_{xyz}, \quad (j = 1, 2, \cdots m), \tag{12}$$

The $m \times m$ inter-inner-products of Eq. (12) lead to the matrix equation as follows

$$X^{(1)} - \sqrt{\Gamma} X_z^{(2)} \sqrt{\Gamma^+} = CC^+, \tag{13}$$

where the $m \times m$ matrixes $X^{(1)}$, $X_z^{(2)}$, and $\Gamma$ are the same as those in Sec. II and $C = [c_{jk}]$. Since $|\Psi_1\rangle, |\Psi_2\rangle, \cdots, |\Psi_m\rangle$ are linearly independent, $X^{(1)}$ is a positive definite matrix. From continuity, for $\gamma_{l+1} = \gamma_{l+2} = \cdots = \gamma_m = 0$ and small enough but positive $\gamma_1, \gamma_2, \cdots, \gamma_l$, the matrix $X^{(1)} - \sqrt{\Gamma} X_z^{(2)} \sqrt{\Gamma^+}$ is positive semidefinite. So the matrix $X^{(1)} - \sqrt{\Gamma} X_z^{(2)} \sqrt{\Gamma^+}$ can be diagonalized by a unitary matrix V as

$$V^+ (X^{(1)} - \sqrt{\Gamma} X_z^{(2)} \sqrt{\Gamma^+}) V = \mathrm{diag}(\chi_1, \chi_2, \cdots, \chi_m), \tag{14}$$

where $\chi_1, \chi_2, \cdots$ and $\chi_m$ are nonnegative numbers. Then one can choose the matrix $C$ in Eq. (13) as

$$C = V \, \mathrm{diag}(\sqrt{\chi_1}, \sqrt{\chi_2}, \cdots, \sqrt{\chi_m}) V^+. \tag{15}$$

Equation (13) is thus satisfied with small enough but positive $\gamma_1, \gamma_2, \cdots, \gamma_l$ and $\gamma_{l+1} = \gamma_{l+2} = \cdots = \gamma_m = 0$. On the contrary, if the choice of $\gamma_1, \gamma_2, \cdots$ and $\gamma_l$ brings about a consequence that the matrix $X^{(1)} - \sqrt{\Gamma} X_z^{(2)} \sqrt{\Gamma^+}$ is negative definite, Equation (13) can not be satisfied because $CC^+$ is a positive semidefinite matrix. We thus get the following theorem.

**Theorem 2** — Linearly dependent states set $S = \{|\Psi_1\rangle, |\Psi_2\rangle, \cdots, |\Psi_n\rangle\}$ can be partly cloned in a probabilistic way if and only if $S = \{S_m, S'_m\} = \{S_l, S'_l\}$, $\gamma_{l+1} = \gamma_{l+2} = \cdots = \gamma_n = 0$ and the $m \times m$ matrix $X^{(1)} - \sqrt{\Gamma} X_z^{(2)} \sqrt{\Gamma^+}$ is positive semidefinite.

The semipositivity of the matrix $X^{(1)} - \sqrt{\Gamma} X_z^{(2)} \sqrt{\Gamma^+}$ leads to a series of inequalities about the cloning efficiencies. The optimal cloning efficiencies can be obtained by solving these inequalities and then taking the maximum over all possible choices of the normalized states $P^{(j)}$. For example, let us study the part PQC of four linearly dependent states set $S = \{|\Psi_1\rangle, |\Psi_2\rangle, |\Psi_3\rangle, |\Psi_4\rangle\}$. From Theorem 1 one knows that $S_l$ must be a nonempty subset. Note that if $S_m = \{|\Psi_1\rangle, |\Psi_2\rangle\}$ and $S_l = \{|\Psi_1\rangle\}$ one can get $|\Psi_2\rangle = |\Psi_3\rangle = |\Psi_4\rangle$ without considering the global phase factor. Moreover, if $S_m = \{|\Psi_1\rangle, |\Psi_2\rangle, |\Psi_3\rangle\}$ and $S_l = \{|\Psi_1\rangle, |\Psi_2\rangle\}$, one can get $|\Psi_3\rangle = |\Psi_4\rangle$. Therefore in the following we assume $S_m = \{|\Psi_1\rangle, |\Psi_2\rangle, |\Psi_3\rangle\}$ and $S_l = \{|\Psi_1\rangle\}$.

In the following, we will deduce the optimal cloning efficiencies by using Theorem 2. Since $S_m = \{|\Psi_1\rangle, |\Psi_2\rangle, |\Psi_3\rangle\}$ and $S_l = \{|\Psi_1\rangle\}$, we have $\gamma_2 = \gamma_3 = \gamma_4 = 0$ and

$$X^{(1)} - \sqrt{\Gamma} X_z^{(2)} \sqrt{\Gamma^+} = \begin{pmatrix} 1 - \gamma_1 & \langle \Psi_1 | \Psi_2 \rangle & \langle \Psi_1 | \Psi_3 \rangle \\ \langle \Psi_2 | \Psi_1 \rangle & 1 & \langle \Psi_2 | \Psi_3 \rangle \\ \langle \Psi_3 | \Psi_1 \rangle & \langle \Psi_3 | \Psi_2 \rangle & 1 \end{pmatrix}. \tag{16}$$

From Theorem 2 one knows that $X^{(1)} - \sqrt{\Gamma} X_z^{(2)} \sqrt{\Gamma^+}$ must be a positive semidefinite matrix. Therefore every principal minor of $X^{(1)} - \sqrt{\Gamma} X_z^{(2)} \sqrt{\Gamma^+}$ should be nonnegative. Then we have the

following inequalities

$$(1-\gamma_1)-|\langle \Psi_1|\Psi_2\rangle|^2 \geq 0, \tag{17}$$

$$(1-\gamma_1)-|\langle \Psi_1|\Psi_3\rangle|^2 \geq 0, \tag{18}$$

$$1-|\langle \Psi_2|\Psi_3\rangle|^2 \geq 0, \tag{19}$$

$$(1-\gamma_1)(1-|\langle \Psi_2|\Psi_3\rangle|^2) + 2\operatorname{Re}(\langle \Psi_1|\Psi_2\rangle\langle \Psi_2|\Psi_3\rangle\langle \Psi_3|\Psi_1\rangle) - |\langle \Psi_1|\Psi_3\rangle|^2 - |\langle \Psi_1|\Psi_2\rangle|^2 \geq 0. \tag{20}$$

Solve the inequalities (17-20) one can get

$$\gamma_1 \leq 1 - \frac{|\langle \Psi_1|\Psi_2\rangle|^2 + |\langle \Psi_1|\Psi_3\rangle|^2 - 2\operatorname{Re}(\langle \Psi_1|\Psi_2\rangle\langle \Psi_2|\Psi_3\rangle\langle \Psi_3|\Psi_1\rangle)}{1-|\langle \Psi_2|\Psi_3\rangle|^2}. \tag{21}$$

## IV  GENERALIZATIONS AND SUMMARY

In this section, we firstly extend the part PQC in Sec. III to the case of $1 \to N$ ($N > 2$) part PQC. Because the quantum state discrimination is similar to PQC, we then consider the part discrimination of linearly dependent states.

The $1 \to 2$ part PQC can be directly generalized to the $1 \to N$ case. The generalization is straightforward, and the proof is omitted. The result is the following theorem.

***Theorem 3*** — Linearly dependent states set $S = \{|\Psi_1\rangle, |\Psi_2\rangle, \cdots, |\Psi_n\rangle\}$ can be partly cloned in a probabilistic way if and only if $S = \{S_m, S'_m\} = \{S_l, S'_l\}$, $\gamma_{l+1} = \gamma_{l+2} = \cdots = \gamma_n = 0$ and the $m \times m$ matrix $X^{(1)} - \sqrt{\Gamma} X_z^{(N)} \sqrt{\Gamma^+}$ is positive semidefinite, where $X_z^{(N)} = \left[\langle \Psi_j|\Psi_k\rangle^N \langle P^{(j)}|P^{(k)}\rangle\right]$.

Now, let us study the part discrimination of linearly dependent states set $S = \{S_m, S'_m\} = \{S_l, S'_l\}$. The discrimination process is expressed as

$$U|\Psi_j\rangle_x |P^{(0)}\rangle_z = \sqrt{\gamma_j}|\Psi^{(j)}\rangle_x |P^{(j)}\rangle_z + \sqrt{1-\gamma_j}|\Phi^{(j)}\rangle_x |P^{(n+1)}\rangle_z, \quad (j=1,2,\cdots n), \tag{22}$$

where $\gamma_j$ is the discrimination efficiency and $|P^{(1)}\rangle$, $|P^{(2)}\rangle$, $\cdots$, $|P^{(n+1)}\rangle$ are orthogonal to each other. In the state discrimination process, a unitary operation $U$ is firstly performed on the composite system $xz$. Then a projection measurement is carried out on the system $z$. The measurement result $|P^{(j)}\rangle$ means that the original state is $|\Psi_j\rangle$; whereas with the measurement result $|P^{(n+1)}\rangle$, the discrimination fails. Similar to the proof of Theorem 2, we get the following theorem.

***Theorem 4*** — Linearly dependent states set $S = \{|\Psi_1\rangle, |\Psi_2\rangle, \cdots, |\Psi_n\rangle\}$ can be partly discriminated if and only if $S = \{S_m, S'_m\} = \{S_l, S'_l\}$, $\gamma_{l+1} = \gamma_{l+2} = \cdots = \gamma_n = 0$ and the $m \times m$ matrix $X^{(1)} - \Gamma$ is positive semidefinite.

Comparing Theorem 3 with Theorem 4 it can be seen that, if $N \to \infty$ we have $X^{(1)} - \sqrt{\Gamma} X_z^{(N)} \sqrt{\Gamma^+} \to X^{(1)} - \Gamma$. Therefore the optimal efficiencies for $1 \to \infty$ part PQC are the same as those for part discrimination. By the same calculations as those in Sec. III we find that the optimal efficiencies of $1 \rightarrow N$ ($N \geq 2$) part PQC and part discrimination of the four linearly dependent states set $S = \{|\Psi_1\rangle, |\Psi_2\rangle, |\Psi_3\rangle, |\Psi_4\rangle\}$ are completely the same. That is, $\gamma_2 = \gamma_3 = \gamma_4 = 0$ and $\gamma_1$ in the inequality (21).

To summarize, in this paper we investigate the PQC of linearly dependent states set. We have shown that if and only if a linearly independent subset can not be expressed as the linear superposition of the other states in the set, the linearly dependent states set can be part cloned. The cloning efficiencies of the states in the subset are positive while the other cloning efficiencies are zero. The optimal possible cloning efficiencies for part PQC are studied. We extend the part PQC to the part discrimination of linearly dependent states set.

## ACKNOWLEDGEMENTS

This work is supported by the Natural Science Foundation of Anhui province (1508085SMA206, 1508085MF136) and the National Natural Science Foundation of China (11374013).